\documentclass[letterpaper,english,aps,letter,english,pre,12pt]{revtex4-1}
\usepackage[T1]{fontenc}
\usepackage[latin9]{inputenc}
\usepackage{xcolor}
\usepackage{pdfcolmk}
\usepackage{babel}
\usepackage{verbatim}
\usepackage{amsmath}
\usepackage{esint}
\PassOptionsToPackage{normalem}{ulem}
\usepackage{ulem}
\usepackage[unicode=true,pdfusetitle,
 bookmarks=true,bookmarksnumbered=false,bookmarksopen=false,
 breaklinks=false,pdfborder={0 0 1},backref=false,colorlinks=false]
 {hyperref}

\makeatletter

\pdfpageheight\paperheight
\pdfpagewidth\paperwidth

\providecolor{lyxadded}{rgb}{1,0,0}
\providecolor{lyxdeleted}{rgb}{0,0,1}

 \@ifundefined{textcolor}{}
 {%
   \definecolor{BLACK}{gray}{0}
   \definecolor{WHITE}{gray}{1}
   \definecolor{RED}{rgb}{1,0,0}
   \definecolor{GREEN}{rgb}{0,1,0}
   \definecolor{BLUE}{rgb}{0,0,1}
   \definecolor{CYAN}{cmyk}{1,0,0,0}
   \definecolor{MAGENTA}{cmyk}{0,1,0,0}
   \definecolor{YELLOW}{cmyk}{0,0,1,0}
 }

\usepackage{ae,aecompl}

\newcommand{\sfrac}[2]{\mathchoice
  {\kern0em\raise.5ex\hbox{\the\scriptfont0 #1}\kern-.15em/
   \kern-.15em\lower.25ex\hbox{\the\scriptfont0 #2}}
  {\kern0em\raise.5ex\hbox{\the\scriptfont0 #1}\kern-.15em/
   \kern-.15em\lower.25ex\hbox{\the\scriptfont0 #2}}
  {\kern0em\raise.5ex\hbox{\the\scriptscriptfont0 #1}\kern-.2em/
   \kern-.15em\lower.25ex\hbox{\the\scriptscriptfont0 #2}}
  {#1\!/#2}}

\makeatother

\begin{document}

\title{Dynamic Density Functional Theory\\
with hydrodynamic interactions and fluctuations}

\author{Aleksandar Donev}

\email{donev@courant.nyu.edu}

\affiliation{Courant Institute of Mathematical Sciences, New York University,
New York, NY 10012}

\author{Eric Vanden-Eijnden}

\email{eve2@courant.nyu.edu}

\affiliation{Courant Institute of Mathematical Sciences, New York University,
New York, NY 10012}
\begin{abstract}
We derive a closed equation for the empirical concentration of colloidal
particles in the presence of both hydrodynamic and direct interactions.
The ensemble average of our functional Langevin equation reproduces
known deterministic Dynamic Density Functional Theory (DDFT) {[}\emph{Rex
and Löwen, Phys. Rev. Lett., 101(14):148302, 2008}{]}, and, at the
same time, it also describes the microscopic fluctuations around the
mean behavior. We suggest separating the ideal (non-interacting) contribution
from additional corrections due to pairwise interactions. We find
that, for an incompressible fluid and in the absence of direct interactions,
the mean concentration follows Fick's law just as for uncorrelated
walkers. At the same time, the nature of the stochastic terms in fluctuating
DDFT is shown to be distinctly different for hydrodynamically-correlated
and uncorrelated walkers. This leads to striking differences in the
behavior of the fluctuations around Fick's law, even in the absence
of pairwise interactions. We connect our own prior work {[}\emph{A.
Donev, T. G. Fai, E. Vanden-Eijnden, J. Stat. Mech., P04004, 2014}{]}
on fluctuating hydrodynamics of diffusion in liquids to the DDFT literature,
and demonstrate that the fluid cannot easily be eliminated from consideration
if one wants to describe the collective diffusion in colloidal suspensions.
\end{abstract}
\maketitle
\global\long\def\V#1{\boldsymbol{#1}}
\global\long\def\M#1{\boldsymbol{#1}}
\global\long\def\Set#1{\mathbb{#1}}

\global\long\def\D#1{\Delta#1}
\global\long\def\d#1{\delta#1}

\global\long\def\norm#1{\left\Vert #1\right\Vert }
\global\long\def\abs#1{\left|#1\right|}

\global\long\def\grad{\M{\nabla}}
\global\long\def\avv#1{\langle#1\rangle}
\global\long\def\av#1{\left\langle #1\right\rangle }

\global\long\def\myhalf{\sfrac{1}{2}}
\global\long\def\mythreehalves{\sfrac{3}{2}}

\global\long\def\sM#1{\M{\mathcal{#1}}}
\global\long\def\dprime{\prime\prime}
\global\long\def\Mob{\sM M}
\global\long\def\J{\sM J}
\global\long\def\S{\sM S}
\global\long\def\L{\sM L}
\global\long\def\R{\M{\mathcal{R}}}

\section{Introduction}

Over the past decade and a half there has been considerable interest
in extending traditional (static) Density Functional Theory (DFT)
for liquids to account for dynamics, leading to Dynamic DFT (DDFT)
\cite{DDFT_Diffusion,SPDE_Diffusion_DDFT,SPDE_Diffusion_Dean,SPDE_Diffusion_Formal,DDFT_Pep}.
Recently, attempts have been made to additionally account for hydrodynamic
interactions (HI) among the particles due to the presence of a viscous
solvent \cite{DDFT_Lowen,DDFT_Hydro_Lowen,DDFT_Pavliotis_PRL,DDFT_Pavliotis,DDFT_Hydro_Inertia},
as necessary when modeling colloidal suspensions. A key feature of
these proposed HI+DDFT theories is that even for the simple case of
non-interacting Brownian particles suspended in a fluid the resulting
equations appear to be different from Fick's law, in contrast to the
case of independent (uncorrelated) Brownian walkers. Here we show
that for the case of non-interacting but hydrodynamically-correlated
Brownian particles one can write down a closed equation for the average
density that is exactly Fick's law, without uncontrolled approximations
such as closures of the BBGKY hierarchy. Furthermore, our equation
includes fluctuations around Fick's law, and sheds light on the controversy
over the difference between deterministic and fluctuating DDFT \cite{DDFT_Diffusion,SPDE_Diffusion_DDFT,SPDE_Diffusion_Dean,SPDE_Diffusion_Formal}.
The derivation presented here follows on our previous work \cite{DiffusionJSTAT}
in which we obtain the same result by using a fluctuating hydrodynamic
formalism. Here we follow an approach originally proposed by Dean
\cite{SPDE_Diffusion_Dean} for the case of uncorrelated Brownian
walkers, and obtain the same equation as derived in \cite{DiffusionJSTAT}
by rather different means. Our work demonstrates that hydrodynamics
is not something that is to be added to Fick's law as non-local correction;
rather, fluctuating hydrodynamics underlies diffusion and gives rise
to Fick's law. This simple yet seemingly frequently missed point is
silently evidenced by the well-known Stokes-Einstein relation, which
relates the diffusion coefficient $\chi\sim k_{B}T/\left(\eta\sigma\right)$
to the temperature $T$, the size of the particles $\sigma$, \emph{and}
the viscosity of the fluid $\eta$.

For consistency, in this paper we use the notation of our prior closely-related
work \cite{DiffusionJSTAT} instead of the notation more common in
the DDFT literature. We start from the overdamped Langevin equations
of Brownian Dynamics (BD), which are often used to model dynamics
of colloidal particles or polymer chains in flow. The Ito equations
of motion for the (correlated) positions of the $N$ particles $\V Q\left(t\right)=\left\{ \V q_{1}\left(t\right),\dots,\V q_{N}\left(t\right)\right\} $
are 
\begin{equation}
d\V Q=-\M M\left(\partial_{\V Q}U\right)dt+\left(2k_{B}T\,\M M\right)^{\frac{1}{2}}d\V{\mathcal{B}}+k_{B}T\left(\partial_{\V Q}\cdot\M M\right)dt,\label{eq:BD_M}
\end{equation}
where $\V{\mathcal{B}}(t)$ is a collection of independent Brownian
motions, $U\left(\V Q\right)$ is a conservative interaction potential,
$\M M\left(\V Q\right)\succeq\M 0$ is a symmetric positive semidefinite
\emph{mobility} block matrix for the collection of particles. The
Fokker-Planck equation (FPE) for the probability density $P\left(\V Q,t\right)$
corresponding to (\ref{eq:BD_M}) is 
\begin{equation}
\frac{\partial P}{\partial t}=\frac{\partial}{\partial\V Q}\cdot\left\{ \M M\left[\frac{\partial U}{\partial\V Q}P+\left(k_{B}T\right)\frac{\partial P}{\partial\V Q}\right]\right\} ,\label{eq:FokkerPlanck}
\end{equation}
and is in detailed-balance (i.e., is time reversible) with respect
to the Gibbs-Boltzmann distribution $\sim\exp\left(-U(\V Q)/k_{B}T\right)$.
A commonly-used model of the mobility matrix, suitable for dilute
suspensions, is the Rotne-Prager pairwise approximation \cite{RotnePrager}.

We will assume here that the mobility is \emph{pairwise }additive,
and that the block of the mobility corresponding to the pair of particles
$i$ and $j$ is a smooth function of \emph{only} the positions of
those particles,
\begin{equation}
\forall\left(i,j\right):\quad\M M_{ij}\left(\V q_{i},\V q_{j}\right)=\frac{\R\left(\V q_{i},\V q_{j}\right)}{k_{B}T},\label{eq:M_R}
\end{equation}
where $\R\left(\V r,\V r^{\prime}\right)$ is a symmetric positive-semidefinite
(SPD) tensor kernel (linear operator mapping vector fields to vector
fields) %
\footnote{Here we adopt the notation of our previous work \cite{DiffusionJSTAT},
except that we have included an additional factor of 2 in $\R$ to
simplify some of the expressions.%
}. The assumption of pairwise additivity is appropriate for low-density
colloidal suspensions, when the typical distance between particles
is significantly larger than the typical size of a particle; at higher
densities complex many-body effects appear which are beyond the scope
of this work. Because we assume that \eqref{eq:M_R} holds even if
$i=j$, the self-diffusion tensor of a particle with position $\V r$
is 
\[
\M{\chi}\left(\V r\right)=\R\left(\V r,\V r\right).
\]
For confined systems, $\R\left(\V r,\V r^{\prime}\right)$ depends
on the positions of the two particles relative to the boundaries and
$\M{\chi}\left(\V r\right)$ may be anisotropic and may depend on
the position; for a translationally-invariant and isotropic system
$\R\left(\V r,\V r^{\prime}\right)\equiv\R\left(\V r-\V r^{\prime}\right)$
and $\M{\chi}\left(\V r\right)=\chi\M I$, where $\chi$ is the self-diffusion
coefficient of the particles. Rex and Löwen \cite{DDFT_Lowen,DDFT_Hydro_Lowen}
assume translational invariance but take a form for the mobility in
which the diagonal elements of the mobility are treated differently
from the off-diagonal ones, 
\begin{eqnarray}
\M M_{ij}\left(\V Q\right) & \equiv\M M_{ij}\left(\V q_{i},\V q_{j}\right) & =\frac{\chi}{k_{B}T}\left[\delta_{ij}\M I+\left(1-\delta_{ij}\right)\M{\omega}_{12}\left(\V q_{i}-\V q_{j}\right)\right],\nonumber \\
 &  & =\frac{\chi}{k_{B}T}\left[\delta_{ij}\left(\M I-\M{\omega}_{12}\left(0\right)\right)+\M{\omega}_{12}\left(\V q_{i}-\V q_{j}\right)\right]\label{eq:mob_Lowen}
\end{eqnarray}
where we have neglected higher-order hydrodynamic corrections and
taken $\omega_{11}=0$, which is appropriate for dilute suspensions.
The Rotne-Prager \cite{RotnePrager} form for $\M{\omega}_{12}\left(\V r\right)$,
which is what Rex and Löwen used in their numerical simuations, can
be written in the form
\begin{equation}
\M{\omega}_{12}\left(\V r\right)=\begin{cases}
{\displaystyle \left(\frac{3\sigma}{4r}+\frac{\sigma^{3}}{2r^{3}}\right)\M I+\left(\frac{3\sigma}{4r}-\frac{3\sigma^{3}}{2r^{3}}\right)\frac{\V r\otimes\V r}{r^{2}}}, & \quad r>2\sigma\\
{\displaystyle \left(1-\frac{9r}{32\sigma}\right)\M I+\left(\frac{3r}{32\sigma}\right)\frac{\V r\otimes\V r}{r^{2}}}, & \quad r\leq2\sigma
\end{cases}\label{eq:RPYTensor}
\end{equation}
where $\sigma$ is the radius of the colloidal particles, and satisfies
the key condition $\M{\omega}_{12}(0)=\M I$. Therefore, the term
involving $\delta_{ij}$ in (\ref{eq:mob_Lowen}) can be deleted and
(\ref{eq:mob_Lowen}) becomes of the form (\ref{eq:M_R}) with 
\[
\R\left(\V q_{i},\V q_{j}\right)\equiv\chi\M{\omega}_{12}\left(\V q_{i}-\V q_{j}\right).
\]
Note that in colloidal suspensions there is typically a hard-core
repulsion that ensures that particles essentially never overlap, which
implies that the behavior of $\M{\omega}_{12}\left(\V r\right)$ for
$r\leq2\sigma$ is not expected to be important. Since the effect
of hydrodynamic interactions is distinct from that of direct interactions,
it is important to also consider the case of an ideal gas in which
the only interparticle interactions are of hydrodynamic origin. Furthermore,
particles can overlap relative to their far-field hydrodynamic radius
for suspensions of soft particles such as star polymer chains \cite{MoriZwanzig_ConstrainedMD}.

In a clever but formal derivation \cite{SPDE_Diffusion_Formal}, Dean
started from the overdamped Langevin equations for a collection of
$N$ interacting Brownian walkers driven by \emph{independent} noise,
i.e., a diagonal mobility matrix $\M M=\left(k_{B}T\right)^{-1}\chi\M I$,
to obtain a closed-form equation for the \emph{empirical} or \emph{fluctuating}
density (concentration) of particles
\begin{equation}
c\left(\V r,t\right)=\sum_{i=1}^{N}\delta\left(\V q_{i}\left(t\right)-\V r\right).\label{eq:c_def}
\end{equation}
For non-interacting particles, this equation can formally be written
as an Ito stochastic partial differential equation (SPDE) \cite{SPDE_Diffusion_Formal},
\begin{equation}
\partial_{t}c=\chi\grad^{2}c+\grad\cdot\left(\sqrt{2\chi c}\,\M{\mathcal{W}}_{c}\right),\label{eq:c_Dean}
\end{equation}
where $\M{\mathcal{W}}_{c}\left(\V r,t\right)$ denotes a spatio-temporal
white-noise vector field. As pointed out in Ref. \cite{SPDE_Diffusion_Formal}
and further elaborated in \cite{SPDE_Diffusion_DDFT}, equation (\ref{eq:c_Dean})
is simply a formal rewriting of (\ref{eq:BD_M}). The only difference
is that the identity of the different particles has been removed by
going from a Lagrangian to an Eulerian description. Importantly, the
solution of (\ref{eq:c_Dean}) should forever remain a sum of delta
functions (whose positions diffuse independently). In fact, the multiplicative
noise SPDE (\ref{eq:c_Dean}) as written has no clear mathematical
interpretation, and neither does the square root of a sum of delta
functions in the noise amplitude. 

Of primary interest in practice are expectation values of the instantaneous
concentration $c\left(\V r,t\right)$, such as the \emph{average}
concentration, which is also the single-particle distribution function
$c^{(1)}\left(\V r,t\right)=\av{c\left(\V r,t\right)}$. Taking an
ensemble average of (\ref{eq:c_Dean}) is trivial because of the linearity
of the deterministic term and the fact that the noise term averages
to zero due to its Ito interpretation, and for the case of non-interacting
particles one simply obtains Fick's law,
\begin{equation}
\partial_{t}c^{(1)}=\chi\grad^{2}c^{(1)}.\label{eq:simple_Fick}
\end{equation}
If direct interactions among the particles are included, one cannot
write a closed form equation and an infinite hierarchy of BBGKY equations
arises; a closure approximation for the higher-order correlation functions
is required, as explained by Marconi and Tarazona \cite{SPDE_Diffusion_Formal}.
It is important to note that (\ref{eq:c_Dean}) and (\ref{eq:simple_Fick})
describe rather different objects; the solution to (\ref{eq:c_Dean})
is a spiky sum of delta functions, and not a smooth average density
or single-particle distribution function as Fick's \emph{deterministic
}law (\ref{eq:simple_Fick}) or traditional (static) DFT describes
\cite{SPDE_Diffusion_Formal,DDFT_Diffusion,SPDE_Diffusion_DDFT}.

As summarized in Ref. \cite{SPDE_Diffusion_DDFT}, Fick's law (\ref{eq:simple_Fick})
can also be obtained by starting from the FPE (\ref{eq:FokkerPlanck})
and integrating over $N-1$ particles to get an equation for the single-particle
distribution function $c^{(1)}\left(\V r,t\right)$. This route was
followed by Rex and Löwen \cite{DDFT_Lowen,DDFT_Hydro_Lowen} in order
to include the effect of hydrodynamic interactions in (\ref{eq:simple_Fick})
and obtain an equation that, at first sight, appears distinctly different
from Fick's law. It is important to note that in order to close the
BBGKY hierarchy some uncontrolled approximations are made in Refs.
\cite{DDFT_Lowen,DDFT_Hydro_Lowen}; we will not require such approximations
until Section \ref{sec:DDFT}. For non-interacting particles, in our
notation, eq. (5) in Ref. \cite{DDFT_Hydro_Lowen} reduces to
\begin{equation}
\partial_{t}c^{(1)}\left(\V r,t\right)=\chi\grad^{2}c^{(1)}\left(\V r,t\right)+\chi\grad\cdot\left(\int\M{\omega}_{12}\left(\V r-\V r^{\prime}\right)\grad^{\prime}c^{(2)}\left(\V r,\V r^{\prime},t\right)\, d\V r^{\prime}\right),\label{eq:c_t_Lowen}
\end{equation}
where $c^{(2)}\left(\V r,\V r^{\prime},t\right)$ is the two-particle
distribution function, and we use$\grad$ to denote gradient with
respect to $\V r$ and $\grad^{\prime}$ with respect to $\V r^{\prime}$.\textcolor{red}{{}
}In this work, we derive an equation for the empirical (fluctuating)
concentration in the presence of hydrodynamic interactions similar
to (\ref{eq:c_Dean}), whose expectation gives (\ref{eq:c_t_Lowen}).
In the absence of direct interactions this equation is given by (\ref{eq:limiting_Ito})
and was previously derived by us using a different approach in Ref.
\cite{DiffusionJSTAT}. In addition to reproducing Fick's law for
the average, (\ref{eq:limiting_Ito}) also describes the long-range
correlated fluctuations around the mean. Here we also include the
effect of direct interactions among the particles.

The first term on the right hand side of (\ref{eq:c_t_Lowen}) is
the familiar \emph{local} Fick's law; but the second term is a \emph{non-local}
diffusion term. It is important to note that the far-field behavior
of the mobility (\ref{eq:RPYTensor}) is given by the scaled Oseen
tensor 
\begin{equation}
\M{\omega}_{12}\left(\V r\right)=\frac{3}{4}\frac{\sigma}{r}\left(\M I+\frac{\V r\otimes\V r}{r^{2}}\right)+O\left(\left(\frac{\sigma}{r}\right)^{3}\right),\label{eq:Oseen}
\end{equation}
which is long-ranged and decays as $r^{-1}$. While it may at first
sight look like $\M{\omega}_{12}\left(\V r\right)$ is small for $r\gg\sigma$,
it should be recalled that the Stokes-Einstein formula $\chi=k_{B}T/\left(6\pi\eta\sigma\right)$
implies that the second term in (\ref{eq:c_t_Lowen}) is independent
\footnote{This is expected since the leading-order hydrodynamic correction comes
from a monopole term (Stokeslet) that corresponds to a \emph{point}
force in a Stokesian fluid.%
} of $\sigma$ since $\chi\M{\omega}_{12}\left(\V r\right)\sim\left(k_{B}T\right)/\left(\eta r\right)$.
The equation of Rex and Löwen (\ref{eq:c_t_Lowen}) therefore implies
that Fick's law needs to be amended with a long-ranged non-local term
even for \emph{dilute} suspensions with \emph{no} direct interactions
among the diffusing particles.

Let us observe, however, that the Rotne-Prager mobility (\ref{eq:RPYTensor})
satisfies an additional key property, $\grad\cdot\M{\omega}_{12}(\V r)=0,$
or more generally,
\begin{equation}
\grad\cdot\R(\V r,\V r^{\prime})=0.\label{eq:div_free}
\end{equation}
This is a direct consequence of the fact that hydrodynamic interactions
(correlations) are mediated by an incompressible fluid \cite{RotnePrager}.
In this case the second term on the right hand side of (\ref{eq:c_t_Lowen})
in fact \emph{vanishes} after a simple integration by parts. Therefore,
Fick's law (\ref{eq:simple_Fick}) for the average concentration remains
valid even in the presence of long-ranged hydrodynamic correlations
among the Brownian walkers.\textcolor{red}{{} }This important physical
implication of (\ref{eq:div_free}) seems to be have been missed in
\cite{DDFT_Lowen,DDFT_Hydro_Lowen} and subsequent works because the
focus in DFT, and therefore DDFT, is almost exclusively on interacting
particles and nonlocal free-energy functionals, and comparatively
little attention seems to have been given to the nonlocal difffusion
aspect of (\ref{eq:c_t_Lowen}). Following the completion of this
work, we learned of an early derivation by Altenberger and Deutch
that showed that, indeed, (\ref{eq:simple_Fick}) holds even in the
presence of hydrodynamic interactions (correlations), see (3.10) in
Ref. \cite{LightScattering_HI}. These authors also made use of and
noted the importance of the divergence-free condition (\ref{eq:div_free}).

It is important to also note another derivation aiming to include
hydrodynamics in DDFT, developed by the authors of Refs. \cite{DDFT_Pavliotis_PRL,DDFT_Pavliotis,DDFT_Hydro_Inertia}.
These authors argue that inertia also needs to be included, and arrive
at an equation that has even more non-local terms than (\ref{eq:c_t_Lowen}).
We believe that these derivations, while careful (even rigorous),
start from an incorrect inertial formulation of the equations of motion
of colloidal particles immersed in fluid. As explained by Hinch \cite{VACF_Langevin}
and later summarized eloquently and clearly by Roux \cite{LangevinDynamics_Theory},
any equation of motion that accounts for inertial effects \emph{must}
include the inertia of the fluid in addition to any excess inertia
of the particles over the fluid they expel. This is because the time
it takes for momentum to diffuse through the liquid, with diffusion
coefficient equal to the kinematic viscosity $\nu=\eta/\rho$ (note
the appearance of the fluid inertia here via the density $\rho$),
is in fact \emph{longer} than inertial time scales. It is therefore
inconsistent to use hydrodynamic friction or mobility functions such
as the Rotne-Prager tensor, which assume steady Stokes flow, i.e.,
infinitely fast momentum diffusion, while including inertia of the
particles explicitly. The only Markovian formulation of the hydrodynamics
of colloidal suspensions that includes \emph{both} hydrodynamics and
thermal fluctuations (Brownian motion) consistently is that of \emph{fluctuating
hydrodynamics} \cite{VACF_FluctHydro,Faxen_FluctuatingHydro,VACF_Langevin}.
Roux starts from the inertial formulation of Hinch \cite{VACF_Langevin}
and derives the overdamped equation of motion (\ref{eq:BD_M}) from
those inertial equations \cite{LangevinDynamics_Theory}. We therefore
consider the overdamped equation (\ref{eq:BD_M}), rather than the
inertial Langevin equations used by a number of authors \cite{DDFT_Pavliotis_PRL,DDFT_Archer,DDFT_Inertial},
as the correct starting point for including hydrodynamics in DDFT.

In our own recent work \cite{DiffusionJSTAT}, we started from a simplified
version of the complete formulation of Hinch \cite{VACF_Langevin}
and Roux \cite{LangevinDynamics_Theory}. In this approximation \cite{SIBM_Brownian,SELM,ForceCoupling_Fluctuations,ISIBM,BrownianBlobs},
the no-slip condition resolved over the surface of the particles is
approximated by an average no-slip condition at the centroid of each
of the particles, and the particles are assumed to be neutrally-buoyant
(but see Ref. \cite{ISIBM} for an extension to account for excess
inertia). Another way to think of the approximation is as a low-order
multipole approximation of the complete hydrodynamics, suitable for
dilute suspensions, and accurate to the \emph{same} order as the Rotne-Prager
far-field approximation \cite{ForceCoupling_Monopole,BrownianBlobs}.
By starting from the simplified fluctuating hydrodynamic formulation
and eliminating the fluid velocity as a fast variable, one can obtain
the overdamped Lagrangian equation (\ref{eq:BD_M}) \cite{SIBM_Brownian,DiffusionJSTAT}.
In Ref. \cite{DiffusionJSTAT} we started from an inertial \emph{Eulerian}
description, i.e., a description involving not the positions of the
individual particles but rather the empirical concentration $c\left(\V r,t\right)$,
and obtained, by adiabatic elimination of the fast fluid velocity,
the overdamped Eulerian Ito SPDE
\begin{align}
\partial_{t}c & =\grad\cdot\left[\M{\chi}\left(\V r\right)\grad c\right]-\V w\cdot\grad c.\label{eq:limiting_Ito}
\end{align}
Here $\V w\left(\V r,t\right)$ is a random velocity field that is
white in time and has a spatial covariance \cite{DiffusionJSTAT},
\begin{align}
\av{\V w\left(\V r,t\right)\otimes\V w\left(\V r^{\prime},t^{\prime}\right)} & =2\R\left(\V r,\V r^{\prime}\right)\delta\left(t-t^{\prime}\right),\label{eq:C_w}
\end{align}
and the incompressibility condition (\ref{eq:div_free}) is assumed
to hold. The ensemble average of (\ref{eq:limiting_Ito}) is nothing
other than Fick's law (\ref{eq:simple_Fick}), and does not include
any non-local diffusion terms because of the incompressibility of
the fluid. It is important to point out that (\ref{eq:limiting_Ito}),
just like (\ref{eq:c_Dean}), describes a spiky sum of delta functions
which are advected by a rapidly-decorrelating random velocity field.
However, (\ref{eq:limiting_Ito}) is distinctly different from (\ref{eq:c_Dean}):
while both equations have multiplicative noise, (\ref{eq:limiting_Ito})
is \emph{linear}, while (\ref{eq:c_Dean}) is \emph{nonlinear}. As
we discuss in more detail in the Conclusions, one can obtain (\ref{eq:c_Dean})
from (\ref{eq:limiting_Ito}) upon taking a suitable (nontrivial)
limit in which $\R\left(\V r,\V r^{\prime}\right)$ becomes highly
localized around $\V r=\V r^{\prime}$.

Here, we connect our prior work to the DDFT literature, by obtaining
the overdamped Eulerian (fluctuating DDFT) equation (\ref{eq:limiting_Ito})
starting from the overdamped Lagrangian equation (\ref{eq:BD_M}),
rather than from the inertial Eulerian formulation as we did in Ref.
\cite{DiffusionJSTAT}. Our argument is essentially a generalization
of that of Dean \cite{SPDE_Diffusion_Dean} and makes specific use
of the hydrodynamic formulation that is hidden in Rotne-Prager-like
approximations to the mobility matrix. As it must, for non-interacting
particles the present calculation gives exactly the same result (\ref{eq:limiting_Ito})
for the empirical concentration and Fick's law (\ref{eq:simple_Fick})
for the average concentration. Furthermore, here we extend our previous
work to account for direct interactions (as opposed to hydrodynamic
interactions) among the diffusing particles. Just as in the work of
Dean \cite{SPDE_Diffusion_Formal}, we obtain a \emph{closed} but
\emph{nonlinear} and \emph{nonlocal }equation\emph{ }for the empirical
(fluctuating) concentration. As expected, in the presence of interactions
it is not possible to write down a closed form for the ensemble-averaged
concentration, and approximate closures are required for two-particle
and three-particle correlation functions \cite{SPDE_Diffusion_Formal,DDFT_Lowen,DDFT_Hydro_Lowen}.

This paper is organized as follows. In Section \ref{sec:DDFT} we
summarize and then derive our key result (\ref{eq:c_fluct_general}),
a fluctuating diffusion equation for a collection of particles interacting
both hydrodynamically and via conservative potentials. In Section
\ref{sec:Diffusion} we discuss coarse-graining (averaging) and the
relation of our work to density functional theory, Fick's macroscopic
law, and fluctuating hydrodynamics, and point to several important
open problems. Finally, we give some conclusions in Section \ref{sec:Conclusions}.

\section{\label{sec:DDFT}Fluctuating DDFT with Hydrodynamic Interactions}

In this section we summarize our main results, and defer the detailed
derivations to Appendix \ref{sec:Derivations}. For completeness,
we will include here a direct interaction among the particles in the
form of a conservative potential that includes an external potential
$U_{1}\left(\V r\right)$ and a pairwise additive potential $U_{2}(\V r,\V r^{\prime})$,
\begin{equation}
U\left(\V Q\right)=\sum_{i=1}^{N}U_{1}(\V q_{i})+\frac{1}{2}\sum_{\substack{i,j=1\\
i\not=j
}
}^{N}U_{2}(\V q_{i},\V q_{j})\label{eq:U_Q}
\end{equation}
where, without loss of generality, we can assume that $U_{2}(\V r,\V r^{\prime})=U_{2}(\V r^{\prime},\V r)$
and $\left[\grad U_{2}(\V r,\V r^{\prime})\right]_{\V r^{\prime}=\V r}=0$.
Note that such an interaction was not included in our prior work \cite{DiffusionJSTAT}.

Here we use (\ref{eq:BD_M},\ref{eq:M_R}) to formally derive a closed-form
SPDE for the empirical concentration (\ref{eq:c_def}). Our calculation
mimics the one performed by Dean for the case of uncorrelated walkers
\cite{SPDE_Diffusion_Dean}. The result of the calculations detailed
in Appendix \ref{sec:Derivations} is the fluctuating hydrodynamic
equation (conservation law) 
\begin{equation}
\begin{aligned}\partial_{t}c(\V r,t) & =-\grad\cdot\left(\V w\left(\V r,t\right)c(\V r,t)\right)+\grad\cdot\left(\M{\chi}(\V r)\grad c(\V r,t)+\V b(\V r,\V r)c(\V r,t)\right)\\
 & +\grad\cdot\left(c(\V r,t)\int\R(\V r,\V r^{\prime})\grad^{\prime}c(\V r^{\prime},t)\, d\V r^{\prime}\right)\\
 & +\left(k_{B}T\right)^{-1}\grad\cdot\left(c(\V r,t)\int\R(\V r,\V r^{\prime})\grad^{\prime}U_{1}(\V r^{\prime})c(\V r^{\prime},t)\, d\V r^{\prime}\right)\\
 & +\left(k_{B}T\right)^{-1}\grad\cdot\left(c(\V r,t)\int\R(\V r,\V r^{\prime})\grad^{\prime}U_{2}(\V r^{\prime},\V r^{\dprime})c(\V r^{\prime},t)c(\V r^{\dprime},t)\, d\V r^{\prime}d\V r^{\dprime}\right),
\end{aligned}
\label{eq:c_fluct_general}
\end{equation}
where $\V b(\V r,\V r^{\prime})=\grad^{\prime}\cdot\R(\V r,\V r^{\prime})$
and $\V w\left(\V r,t\right)$ is a random velocity field with covariance
(\ref{eq:C_w}), see (\ref{eq:6c}) for a derivation of the stochastic
term in the Ito convention and (\ref{eq:6cS}) for the Stratonovich
interpretation. Compare (\ref{eq:c_fluct_general}) to the equation
obtained by following the same procedure for the case of uncorrelated
particles, $\M M_{ij}=\delta_{ij}\left(k_{B}T\right)^{-1}\M{\chi}\left(\V q_{i}\right)$,
\begin{eqnarray}
\partial_{t}c(\V r,t) & = & \grad\cdot\left(\left(2\M{\chi}\left(\V r\right)c(\V r,t)\right)^{\frac{1}{2}}\,\M{\mathcal{W}}_{c}\right)+\grad\cdot\left(\M{\chi}(\V r)\grad c(\V r,t)\right)\label{eq:c_Dean_interacting}\\
 & + & \left(k_{B}T\right)^{-1}\grad\cdot\left(\M{\chi}(\V r)c(\V r,t)\grad U_{1}(\V r)\right)\nonumber \\
 & + & \left(k_{B}T\right)^{-1}\grad\cdot\left(\M{\chi}(\V r)c(\V r,t)\int\grad^{\prime}U_{2}(\V r,\V r^{\prime})c(\V r^{\prime},t)\, d\V r^{\prime}\right)\nonumber 
\end{eqnarray}
which is a slight generalization of Eq. (17) in \cite{SPDE_Diffusion_Dean}
to account for the one-particle potential and the possible anisotropy
and spatial dependence of the diffusion tensor $\M{\chi}(\V r)$. 

Ensemble averaging (\ref{eq:c_fluct_general}) gives the first member
of a BBGKY-like hierarchy of equations for the single-particle distribution
function,
\begin{eqnarray}
\partial_{t}c^{(1)}\left(\V r,t\right) & = & \grad\cdot\left(\M{\chi}(\V r)\grad c^{(1)}\left(\V r,t\right)\right)+\grad\cdot\left(\int\R\left(\V r,\V r^{\prime}\right)\grad^{\prime}c^{(2)}\left(\V r,\V r^{\prime},t\right)\, d\V r^{\prime}\right)\nonumber \\
 & + & \left(k_{B}T\right)^{-1}\grad\cdot\left(\M{\chi}(\V r)\grad U_{1}\left(\V r\right)c^{(1)}\left(\V r,t\right)+\int\R\left(\V r,\V r^{\prime}\right)\grad^{\prime}U_{1}\left(\V r^{\prime}\right)c^{(2)}\left(\V r,\V r^{\prime},t\right)\, d\V r^{\prime}\right)\nonumber \\
 & + & \left(k_{B}T\right)^{-1}\grad\cdot\left(\int\left(\M{\chi}(\V r)\grad U_{2}\left(\V r,\V r^{\prime}\right)+\R\left(\V r,\V r^{\prime}\right)\,\grad^{\prime}U_{2}\left(\V r,\V r^{\prime}\right)\right)\, c^{(2)}\left(\V r,\V r^{\prime},t\right)\, d\V r^{\prime}\right)\nonumber \\
 & + & \left(k_{B}T\right)^{-1}\grad\cdot\left(\int\R\left(\V r,\V r^{\prime}\right)\, c^{(3)}\left(\V r,\V r^{\prime},\V r^{\prime\prime},t\right)\,\grad^{\prime}U_{2}\left(\V r^{\prime},\V r^{\prime\prime}\right)\, d\V r^{\prime\prime}d\V r^{\prime}\right),\label{eq:c_av_general}
\end{eqnarray}
which is a slight generalization of equation (5) in Ref. \cite{DDFT_Lowen,DDFT_Hydro_Lowen}
with $\omega_{11}=0$. Here $c^{(3)}(\V r,\V r^{\prime},\V r^{\dprime},t)$
denotes the three-particle correlation function. We note that the
term involving $c^{(3)}$ is missing in (4.4) in Ref. \cite{LightScattering_HI},
as well as (3.1) in Ref. \cite{CollectiveDiffusion_1}, apparently
because of an additional low-density approximation in the spirit of
kinetic theory.

When the incompressibility condition (\ref{eq:div_free}) is satisfied,
Eqs. (\ref{eq:c_fluct_general}) and (\ref{eq:c_av_general}) simplify
in a key way; as also observed in Ref. \cite{LightScattering_HI},
after an integration by parts the nonlocal diffusion term on the second
line of (\ref{eq:c_fluct_general}) and the second term on the right
hand side in the first line of (\ref{eq:c_av_general}) disappear,
see (\ref{eq:6inc}) in the Ito convention and (\ref{eq:6incS}) for
the Stratonovich interpretation %
\footnote{Note that for incompressible $\V w$ we have $\grad\cdot\left(\V wc\right)=\V w\cdot\grad c$.%
}. Therefore, in the absence of interactions the fluctuating DDFT equation
(\ref{eq:c_fluct_general}) reduces to (\ref{eq:limiting_Ito}) and
the mean follows the local Fickian diffusion equation (\ref{eq:simple_Fick}),
\emph{even} in the presence of hydrodynamic correlations among the
particles. This important physical consequence of incompressibility
was not observed by Rex and Löwen \cite{DDFT_Lowen,DDFT_Hydro_Lowen},
and this omission may have lead some readers to the wrong conclusion
that hydrodynamic interactions lead to nonlocal corrections to Fick's
law for the mean.

Although not apparent at first sight, (\ref{eq:c_fluct_general})
has the same structure of an overdamped Langevin equation as does
(\ref{eq:BD_M}), namely, we can formally write it in the compact
notation \cite{OttingerBook}
\begin{equation}
\partial_{t}c=-\M{\mathcal{M}}\left[c(\cdot,t)\right]\frac{\d H}{\d c\left(\cdot,t\right)}+\left(2k_{B}T\,\M{\mathcal{M}}\left[c(\cdot,t)\right]\right)^{\frac{1}{2}}\V{\mathcal{W}}_{c}(\cdot,t)+k_{B}T\left(\frac{\delta}{\d c\left(\cdot,t\right)}\cdot\M{\mathcal{M}}\left[c(\cdot,t)\right]\right),\label{eq:c_Lang_compact}
\end{equation}
where the mobility $\M{\mathcal{M}}\left[c(\cdot)\right]$ is a positive-semidefinite
linear operator that is a functional of the function of position $c$,
denoted here by the notation $\left[c(\cdot)\right]$, and products
imply a contraction over spatial position. More precisely, 
\begin{eqnarray}
\partial_{t}c(\V r,t) & = & -\int d\V r^{\prime}\mathcal{M}\left[c(\cdot,t);\,\V r,\V r^{\prime}\right]\frac{\d H}{\d c\left(\V r^{\prime},t\right)}\label{eq:c_Langevin}\\
 & + & \left(2k_{B}T\right)^{\frac{1}{2}}\int d\V r^{\prime}\,\mathcal{\M{\mathcal{M}}}^{\frac{1}{2}}\left[c(\cdot,t);\,\V r,\V r^{\prime}\right]\V{\mathcal{W}}_{c}(\V r^{\prime},t)\nonumber \\
 & + & \left(k_{B}T\right)\int d\V r^{\prime}\,\left(\frac{\delta\mathcal{M}\left[c(\cdot,t);\,\V r,\V r^{\prime}\right]}{\d c\left(\V r^{\prime},t\right)}\right),\nonumber 
\end{eqnarray}
where the mobility $\mathcal{M}\left[c(\cdot)\right]\left(\V r,\V r^{\prime}\right)\equiv\mathcal{M}\left[c(\cdot);\,\V r,\V r^{\prime}\right]$
is defined by its action on a scalar field $f(\V r)$,
\[
\int d\V r^{\prime}\mathcal{M}\left[c(\cdot);\,\V r,\V r^{\prime}\right]\, f(\V r^{\prime})\equiv-\left(k_{B}T\right)^{-1}\,\grad\cdot\left(c(\V r)\int\R\left(\V r,\V r^{\prime}\right)c(\V r^{\prime})\grad^{\prime}f(\V r^{\prime})\, d\V r^{\prime}\right).
\]
Here $H\left[c(\V r)\right]$ is an energy functional consisting of
an ideal and an excess (potential) contribution,
\[
H\left[c\left(\cdot\right)\right]=H_{\text{id}}\left[c\left(\cdot\right)\right]+H_{\text{exc}}\left[c\left(\cdot\right)\right]=H_{\text{id}}\left[c\left(\cdot\right)\right]+\int U_{1}(\V r)c(\V r)d\V r+\frac{1}{2}\int U_{2}(\V r,\V r^{\prime})c(\V r)c(\V r^{\prime})\, d\V rd\V r^{\prime},
\]
where the ideal gas energy functional is
\[
H_{\text{id}}\left[c\left(\cdot\right)\right]=k_{B}T\int c\left(\V r\right)\left(\ln\left(\Lambda^{3}c\left(\V r\right)\right)-1\right)\, d\V r,
\]
$\Lambda$ is a constant (e.g., the thermal de Broglie wavelength),
and $H_{\text{exc}}$ is the excess free energy functional. It is
important to note that when incompressibility condition (\ref{eq:div_free})
holds, we can remove the ideal contribution from $H$ and define $H\equiv H_{\text{exc}}$
without affecting (\ref{eq:c_Langevin}), because
\[
\int\R\left(\V r,\V r^{\prime}\right)c(\V r^{\prime})\grad^{\prime}\left(\frac{\d H_{\text{id}}}{\d c\left(\V r^{\prime}\right)}\right)\, d\V r^{\prime}=\int\R\left(\V r,\V r^{\prime}\right)\grad^{\prime}c(\V r^{\prime})\, d\V r^{\prime}=0.
\]
Also note that in the case of independent (uncorrelated) particles,
(\ref{eq:c_Dean_interacting}) can be written as a functional Langevin
equation (\ref{eq:c_Langevin}) with the same free-energy functional
but with a different mobility operator $\M{\mathcal{M}}_{\text{ind}}$,
defined by its action on a scalar field $f(\V r)$, 
\[
\int d\V r^{\prime}\mathcal{M}_{\text{ind}}\left[c(\cdot);\,\V r,\V r^{\prime}\right]\, f(\V r^{\prime})\equiv-\left(k_{B}T\right)^{-1}\,\grad\cdot\left(\M{\chi}(\V r)c(\V r)\grad f(\V r)\right).
\]

The kinetic form \cite{KineticStochasticIntegral_Ottinger} of the
(formal) functional FPE associated with (\ref{eq:c_Langevin}) implies
that the equilibrium distribution associated with (\ref{eq:c_fluct_general}),
assumed to be unique, is the formal Gibbs-Boltzmann distribution
\begin{equation}
P\left[c(\cdot)\right]=Z^{-1}\exp\left(-\frac{H\left[c\left(\cdot\right)\right]}{k_{B}T}\right),\label{eq:GB_functional}
\end{equation}
which is the field representation of the equilibrium distribution
$\exp\left(-U(\V Q)/k_{B}T\right)$ associated with the particle description
(\ref{eq:BD_M}). In the incompressible case, uniqueness of the Gibbs-Boltzmann
distribution can be ensured by adding a small multiple of the identity
(so-called bare diffusion \cite{DiffusionJSTAT}) to the mobility
matrix $\M M$, that is, by adding a small multiple of $\M{\mathcal{M}}_{\text{ind}}$
to the mobility operator $\M{\mathcal{M}}$.

\section{\label{sec:Diffusion}Coarse-Graining}

As noted by Marconi and Tarazona \cite{SPDE_Diffusion_Formal}, (\ref{eq:c_fluct_general})
contains the same physical content as (\ref{eq:BD_M}) because we
have not performed any \emph{coarse graining} or averaging, and have
not lost any information except the particle numbering. Nevertheless,
(\ref{eq:c_fluct_general}) is an informative nontrivial rewriting
of (\ref{eq:BD_M}) that can be used to perform additional \emph{coarse-graining}
and attempt to describe the behavior of collective diffusion in colloidal
suspensions at a spectrum of length (and thus also time) scales, going
from a microscopic scale $\xi$ to macroscopic scales. Here we discuss
three distinct types of coarse-graining one can perform on (\ref{eq:c_fluct_general}):
an ensemble average over the realizations of the noise, an average
over an ensemble of initial conditions, and spatial averaging over
a large number of particles \cite{CoarseGraining_Pep}. Spatial averaging
is of great interest in practice since colloidal suspensions are typically
observed at mesoscopic scales larger than the size of individual particles.
For example, in typical experiments such as light scattering from
colloidal suspensions, concentration fluctuations are averaged over
a region containing many particles (e.g., the thickness of the sample).

One of the simplest, though by no means the only \cite{DiscreteDiffusion_Espanol},
ways to approach such spatial coarse graining is to define a smoothed
empirical concentration that averages over particles in a physical
region of typical size $\xi$ (see Section 4 in Ref. \cite{SPDE_Diffusion_DDFT}
and Section IV in Ref. \cite{DiffusionJSTAT}),
\begin{equation}
c_{\xi}\left(\V r,t\right)=\sum_{i=1}^{N}\delta_{\xi}\left(\V q_{i}\left(t\right)-\V r\right),\label{eq:c_def_CG}
\end{equation}
where $\delta_{\xi}$ is a smoothing kernel with support $\sim\xi$
that converges to a delta function as $\xi\rightarrow0$ (e.g., a
Gaussian with standard deviation $\xi$). For $\xi$ smaller than
the typical particle size or interparticle distance, we have little
to no coarse-graining and detailed microstructural information (e.g.,
layering in a hard-core fluid) is encoded in $c_{\xi}$. For $\xi$
much larger than some characteristic correlation length (e.g., decay
length of the pair correlation function), microstructural information
will no longer be encoded in $c_{\xi}$, although fluctuations in
$c_{\xi}$ may still be non-negligible. Ultimately, for very large
$\xi$ we expect $c_{\xi}$ to become \emph{macroscopic} with negligible
fluctuations, although it is not \emph{a priori} obvious how large
$\xi$ needs to be for this to become the case.

\subsection{Ensemble Averaging}

For simplicity, and in order to facilitate a direct comparison with
prior work by others, in this section we will assume there is no external
potential, $U_{1}=0$. We will also assume an isotropic homogeneous
(translationally- and rotationally-invariant) system,

\[
\R(\V r,\V r^{\prime})\equiv\R(\V r-\V r^{\prime})\quad\text{and}\quad\M{\chi}(\V r)\equiv\chi\M I.
\]
Furthermore, we will assume that the incompressibility condition (\ref{eq:div_free})
holds, which we again emphasize is true for the Rotne-Prager mobility.

Direct ensemble averaging of the functional Langevin equation (\ref{eq:c_fluct_general})
gives
\begin{eqnarray}
\partial_{t}c^{(1)}(\V r,t) & = & -\int d\V r^{\prime}\av{\mathcal{M}\left[c(\cdot,t);\,\V r,\V r^{\prime}\right]\frac{\d H_{\text{exc}}}{\d c\left(\V r^{\prime},t\right)}}\label{eq:ens_av}\\
 & + & \left(k_{B}T\right)\int d\V r^{\prime}\,\av{\frac{\delta\mathcal{M}\left[c(\cdot,t);\,\V r,\V r^{\prime}\right]}{\d c\left(\V r^{\prime},t\right)}},\nonumber 
\end{eqnarray}
where we used the fact that for incompressible $\R$ we can replace
$H$ by $H_{\text{exc}}$, and the fact that in the Ito interpretation
the stochastic term vanishes in expectation. As derived more carefully
in Appendix A of our prior work \cite{DiffusionJSTAT}, the thermal
or stochastic drift term on the second line of (\ref{eq:ens_av})
can be averaged explicitly due to linearity, and leads to the first
term on the right hand side of (\ref{eq:c_av_general}). This demonstrates
that Fickian diffusion is already included in the hydrodynamic correlation
tensor $\R$, as evidenced by the Stokes-Einstein-like relation $\chi\M I=\R\left(0\right)$.
It also shows that all of the terms in the second, third and fourth
lines of (\ref{eq:c_av_general}) come from the closure of the term
$\av{-\M{\mathcal{M}}\,\d H_{\text{exc}}/\d c}$. Recall that the
second term on the first line of (\ref{eq:c_av_general}) disappears
for incompressible $\R$.

In order to make (\ref{eq:c_av_general}) useful in practice, some
closure approximation for the two-particle correlation function is
required, and it is here that equilibrium statistical mechanical quantities
such as free energy functionals enter in the calculations, as first
discussed by Marconi and Tarazona \cite{SPDE_Diffusion_Formal} in
the absence of hydrodynamic correlations and then generalized by Rex
and Löwen \cite{DDFT_Lowen,DDFT_Hydro_Lowen} to account for hydrodynamics.
Namely, by assuming that the higher-order correlation functions can
be approximated by those of the equilibrium system kept at the same
density profile by an external potential, system (\ref{eq:c_av_general})
can be \emph{approximated} with (c.f. (14) in \cite{DDFT_Hydro_Lowen})
\begin{eqnarray}
\partial_{t}c^{(1)}\left(\V r,t\right) & = & \left(k_{B}T\right)^{-1}\chi\grad\cdot\left(c^{(1)}\left(\V r,t\right)\grad\frac{\d F}{\d{c^{(1)}}\left(\V r,t\right)}\right)\label{eq:DDFT_Lowen}\\
 & + & \left(k_{B}T\right)^{-1}\grad\cdot\left(\int\R\left(\V r-\V r^{\prime}\right)c^{(2)}\left(\V r,\V r^{\prime},t\right)\grad^{\prime}\frac{\d F}{\d{c^{(1)}}\left(\V r^{\prime},t\right)}\, d\V r^{\prime}\right),\nonumber 
\end{eqnarray}
where $F\left[c^{(1)}\left(\cdot\right)\right]$ is the \emph{equilibrium}
density functional familiar from static DFT, which is only explictly
known for the ideal-gas, see the discussion around (\ref{eq:DDFT_Fick}).
This microscopic equilibrium density functional captures microstructural
information about the colloidal system at scales comparable to the
size of the colloidal particles. Español and Löwen \cite{DDFT_Pep}
explain how to connect the equilibrium free-energy functional with
a non-Markovian non-local equation for $c^{(1)}$ \emph{without} making
approximations; after making a Markovian (separation of time scales)
approximation they obtain a non-local diffusion equation (c.f. (32)
in Ref. \cite{DDFT_Pep}), and after a further approximation of the
diffusion kernel they obtain the equation of Marconi and Tarazona.
Note that in the presence of hydrodynamic correlations the second
line of (\ref{eq:DDFT_Lowen}) involves $c^{(2)}$, which makes the
equation unclosed and therefore not yet useful in practice without
a further closure approximation for $c^{(2)}\left(\V r,\V r^{\prime},t\right)$.
Rex and Löwen \cite{DDFT_Lowen,DDFT_Hydro_Lowen} suggest such an
approximation in terms of the equilibrium pair correlation function.

It is important to note that, in general, the free-energy functional
$F$ (defined on a space of functions) that enters in the equation
for the ensemble average is \emph{different} from the energy functional
$H$ (formally defined on a space of distributions) that enters in
the functional Langevin equation (\ref{eq:c_Langevin}). In fact,
a precise thermodynamic definition can be given to the classical DDFT
functional $F\left[c^{(1)}\left(\cdot\right)\right]$ as an expectation
value over a Gibbs-Boltzmann distribution related to (\ref{eq:GB_functional}).
However, for noninteracting particles (an ideal gas) $F$ and $H$
have formally the same functional form,
\[
F=F_{\text{id}}=H_{\text{id}}.
\]
Equation (\ref{eq:DDFT_Lowen}) as written contains a long-ranged
nonlocal diffusion term on the second line, which is there even when
there are no direct interactions. For an ideal gas, the flux in the
parenthesis on the second line of (\ref{eq:DDFT_Lowen}) becomes
\[
\int\R\left(\V r-\V r^{\prime}\right)\frac{c^{(2)}\left(\V r,\V r^{\prime},t\right)}{c^{(1)}\left(\V r^{\prime},t\right)}\grad^{\prime}c^{(1)}\left(\V r^{\prime},t\right)\, d\V r^{\prime},
\]
which is still not closed. For an ideal gas, the closure for the two-particle
correlation function that Rex and Löwen \cite{DDFT_Lowen,DDFT_Hydro_Lowen}
suggest becomes
\[
c^{(2)}\left(\V r,\V r^{\prime},t\right)\approx c^{(1)}\left(\V r,t\right)c^{(1)}\left(\V r^{\prime},t\right).
\]
After also making this approximation we can write the second line
of (\ref{eq:DDFT_Lowen}) in the form 
\[
c^{(1)}\left(\V r,t\right)\int\R\left(\V r-\V r^{\prime}\right)\grad^{\prime}c^{(1)}\left(\V r^{\prime},t\right)\, d\V r^{\prime},
\]
which vanishes after an integration by parts due to the incompressibility
condition (\ref{eq:div_free}).

The above considerations for an ideal gas suggest that (\ref{eq:DDFT_Lowen})
should be written in a form that separates the ideal from the non-ideal
contributions,
\begin{eqnarray}
\partial_{t}c^{(1)}\left(\V r,t\right) & = & \chi\grad^{2}c^{(1)}\left(\V r,t\right)+\left(k_{B}T\right)^{-1}\chi\grad\cdot\left(c^{(1)}\left(\V r,t\right)\grad\frac{\d{F_{\text{exc}}}}{\d{c^{(1)}}\left(\V r,t\right)}\right)\nonumber \\
 & + & \left(k_{B}T\right)^{-1}\grad\cdot\left(\int\R\left(\V r-\V r^{\prime}\right)c^{(2)}\left(\V r,\V r^{\prime},t\right)\grad^{\prime}\frac{\d{F_{\text{exc}}}}{\d{c^{(1)}}\left(\V r^{\prime},t\right)}\, d\V r^{\prime}\right).\label{eq:DDFT_Fick}
\end{eqnarray}
where $F_{\text{exc}}$ is the \emph{excess} (over the ideal gas)
free-energy functional. The first line is the equation obtained for
uncorrelated walkers by Marconi and Tarazona \cite{SPDE_Diffusion_Formal}.
In the last line of (\ref{eq:DDFT_Fick}), $\R$ is long-ranged but
one expects that the remainder of the integrand is short-ranged far
from phase transitions in some sense \cite{CollectiveDiffusion_1}
and therefore the result will be nonlocal only over scales that represents
that typical correlation length in the microstructure of the system.
Making this more precise requires some further approximations and
is beyond the scope of this work. It is interesting to note that the
first line in (\ref{eq:DDFT_Fick}) can be written in functional notation
as
\[
-\int d\V r^{\prime}\,\mathcal{M}_{\text{ind}}\left[c^{(1)}(\cdot,t);\,\V r,\V r^{\prime}\right]\,\frac{\d F}{\d c^{(1)}\left(\V r^{\prime},t\right)},
\]
which, surprisingly, involves $\mathcal{M}_{\text{ind}}$ even though
$\mathcal{M}_{\text{ind}}$ does not appear in the original dynamics.
Further work is necessary to explore how well closures such as (\ref{eq:DDFT_Fick})
describe collective diffusion in both confined and unconfined dilute
and semi-dilute colloidal suspensions.

\subsection{Averaging over initial conditions}

As written, the fluctuating DDFT equation (\ref{eq:c_fluct_general})
is a nonlinear non-local SPDE that appears of little practical utility;
solving it is no easier than solving (\ref{eq:BD_M}) using Brownian
Dynamics \cite{BrownianBlobs}. This is so even in the absence of
direct interactions because of the nonlocal diffusive flux term $c(\V r,t)\int\R(\V r,\V r^{\prime})\grad^{\prime}c(\V r^{\prime},t)\, d\V r^{\prime}$.
However, an important observation, previously missed, is that the
incompressibility of the fluid mediating the hydrodynamic correlations
implies that the correlation tensor is divergence free. This implies
that the nonlocal diffusive flux term vanishes, and therefore, in
the absence of direct interactions the fluctuating DDFT equation is
the \emph{linear} and \emph{local} stochastic advection-diffusion
equation (\ref{eq:limiting_Ito}). 

It is important to emphasize that (\ref{eq:limiting_Ito}) is mathematically
well-behaved and \emph{does} have utility beyond that of formal equations
such as (\ref{eq:c_Dean_interacting}) because it can be averaged
over initial conditions (rather than over realizations of the noise)
\cite{DiffusionJSTAT}. Specifically, let us assume that the initial
positions of the particles are uniformly sampled from an equilibrium
ensemble \emph{constrained} to have a specified mean $c_{0}\left(\V r,t\right)$
via a suitable external or chemical potential \cite{SPDE_Diffusion_Formal,DDFT_Hydro_Lowen}.
For noninteracting walkers, this simply amounts to choosing the initial
particle positions independently from a probability distribution $\sim c_{0}\left(\V r,t\right)$.
Because of the linearity of (\ref{eq:limiting_Ito}) we can trivially
average it over this ensemble of initial conditions; the equation
remains the same but now the initial condition is the \emph{smooth}
$c\left(\V r,0\right)=c_{0}$ rather than a spiky sum of delta functions.
This is useful if one wants to describe particular instances (realizations)
of the dynamics starting from a random configuration of particles.
For example, consider a fluorescence recovery after photobleaching
(FRAP) experiment \cite{FRAP_Review} in which a random subset of
the particles uniformly distributed \emph{below} a given plane are
fluorescently labeled at $t=0$ and then allowed to diffuse freely.
This can be modeled by solving (\ref{eq:BD_M}) for a finite collection
of particles, but, equivalently, one can solve (using computational
fluid dynamics techniques) the Eulerian equation (\ref{eq:limiting_Ito})
with $c\left(\V r,0\right)=\text{const}.$ above the given plane and
$c\left(\V r,0\right)=0$ below it, to obtain the probability $\sim c\left(\V r,t\right)$
of finding a particle at position $\V r$ for a \emph{specific} instance
of the noise $\V w\left(\V r,t\right)$. More general smooth initial
conditions are also possible, e.g., a Gaussian profile corresponding
to a nonuniform laser beam intensity in a FRAP experiment.

Because of its nonlinearity, averaging (\ref{eq:c_fluct_general})
over initial conditions is nontrivial and requires further approximations
that are beyond the scope of this work. We believe such averaging
could lead to descriptions that describe collective diffusion at \emph{all}
scales, from the microscopic to the macroscopic, in a manner more
suitable for numerical approximations than (\ref{eq:BD_M}).

\subsection{Spatial Averaging}

It is important to contrast the fluctuating diffusion (\ref{eq:DDFT_Fick})
that describes the microscopic dynamics to the equation obtained by
considering a \emph{macroscopic} limit and coarse-graining the concentration
over many particles, rather than over realizations of the noise. The
literature on the subject is large \cite{CollectiveDiffusion_3,CollectiveDiffusion_2,CollectiveDiffusion_1,LightScattering_HI,CollectiveDiffusion_4}
and we make no attempt to review it here, rather, we summarize some
key results. Let us denote with $\bar{c}\left(\V r,t\right)\approx c_{\xi}\left(\V r,t\right)$
the macroscopic concentration, which, roughly speaking, can be thought
of as $c\left(\V r,t\right)$ averaged over a region of macroscopic
size $\xi$ (i.e., a region containing many particles and typical
size much larger than the interaction range of the pairwise potential).
A precise mathematical definition is possible by suitable rescaling
of space and time, see Refs. \cite{LebowitzHydroReview,LLN_InteractingBrownian,CLT_InteractingBrownian,CLT_InteractingBrownian_2};
equivalently, one can consider the Fourier transform of $c\left(\V r,t\right)$
in the limit of small wavenumbers. It has been demonstrated rigorously
\cite{LLN_InteractingBrownian} that for uncorrelated walkers interacting
with short-ranged potentials the macroscopic concentration obeys a
nonlinear but local Fick's law \cite{CollectiveDiffusion_3}
\[
\partial_{t}\bar{c}=\chi\grad^{2}\Pi(\bar{c})=\chi\grad\cdot\left(\frac{d\Pi(\bar{c})}{d\bar{c}}\grad\bar{c}\right).
\]
Here $\Pi(\bar{c})$ is the osmotic pressure of the suspension at
thermodynamic (local) equilibrium with uniform concentration $\bar{c}$
(for an ideal gas $\Pi(\bar{c})=\bar{c}\, k_{B}T$), $\Pi(\bar{c})=\bar{c}\left(df/d\bar{c}\right)-f$,
where $f(\bar{c})$ is the thermodynamic \emph{equilibrium} free-energy
density of a \emph{macroscopic} system with uniform density $\bar{c}$.

We are, however, not aware of any mathematical techniques that can
be used to rigorously justify Fick's law in the presence of long-ranged
hydrodynamic correlations. Felderhof \cite{CollectiveDiffusion_1}
argues that from a variant of (\ref{eq:c_av_general}) one can obtain
Fick's law with a diffusion coefficient that depends on concentration
and gives a low-density expansion of the collective diffusion coefficient
(c.f. (4.24) in \cite{CollectiveDiffusion_1}) that matches the one
obtained by Batchelor \cite{CollectiveDiffusion_4} using Einstein's
formula. It is important to point out that at later stages of his
argument Felderhof makes key use of the divergence-free nature of
the hydrodynamic correlations %
\footnote{This part of the derivation of Felderhof inspired the rewriting (\ref{eq:DDFT_Fick}).%
}, which he also emphasizes follows from the incompressibility of the
fluid (c.f. (4.13) in \cite{CollectiveDiffusion_1}). While Felderhof
and other authors in the physics literature write Fick's law as an
equation for $c^{(1)}$ it is clear from the derivations that an assumption
is being made that $c^{(1)}$ varies little and slowly in space. It
is important to remember that $c^{(1)}\left(\V r,t\right)$ and $\bar{c}\left(\V r,t\right)$
are different objects, although one expects that in cases where $c^{(1)}$
varies slowly in space the two are closely related since ensemble
and spatial averaging are expected to commute.

Of particular interest is to understand collective diffusion over
the broad-spectrum of \emph{mesoscopic} length-scales, i.e., scales
that are larger than $\sigma$, where $\sigma$ is a typical microscopic
length, but not so large that the hydrodynamic limit applies. For
non-interacting uncorrelated walkers, the ensemble-averaged concentration
follows the \emph{same} diffusion equation (Fick's law) with the same
diffusion coefficient at \emph{all} scales, as seen from the linearity
of (\ref{eq:simple_Fick}). We demonstrated here that the same holds
even in the presence of hydrodynamic correlations among the particles.
Direct interactions appear to, however, complicate the picture and
lead to non-local nonlinear terms like those in (\ref{eq:DDFT_Fick}),
and we do not know of any rigorous results in the mesoscopic regime.
Non-equilibrium thermodynamics \cite{OttingerBook} and the theory
of coarse-graining \cite{CoarseGraining_Pep} provide guidance on
the structure of the resulting equations but not their explicit form.

In principle, an equation for the coarse-grained concentration (\ref{eq:c_def_CG})
can be carried out by convolving (filtering) the right hand side of
(\ref{eq:c_fluct_general}) with the kernel $\delta_{\xi}$. In general
this leads to an unclosed equation and further approximations are
required. Once again the special case of an ideal gas is much simpler
to tackle because (\ref{eq:c_fluct_general}) becomes the linear (\ref{eq:limiting_Ito}).
In Ref. \cite{DiffusionJSTAT} we proposed how to carry out spatial
coarse-graining by performing a \emph{partial} ensemble average of
(\ref{eq:limiting_Ito}) over fluctuations of the random velocity
field $\V w$ below the coarse-graining scale. Our argument, however,
closely relied on the linearity of (\ref{eq:limiting_Ito}) and therefore
only applies when there are incompressible hydrodynamic correlations
but no direct interactions among the particles. The general conclusion
of our work and other related works in the literature is that coarse-graining
leads to effective dissipation (entropy production) with transport
coefficients that must be \emph{renormalized} in a way that takes
into account the mesoscopic observation scale. The same undoubtly
holds for any ``free energy functional'' that may appear in the
mesoscopic equations. Carrying out such a renormalization of (\ref{eq:c_fluct_general})
remains a difficult but important challenge for the future.

\section{\label{sec:Conclusions}Conclusions}

Hydrodynamics plays an important role in colloidal suspensions and
must be included in DDFT theories. Momentum transport in the fluid
leads to hydrodynamic correlations among the diffusing particles and
has important consequences for the collective diffusion not seen if
one looks at the self-diffusion of a single particle in suspension.
Starting from \eqref{eq:M_R} as a model of these hydrodynamic correlations,
we obtained a closed equation (\ref{eq:c_fluct_general}) for the
instantaneous, fluctuating, or empirical concentration, the ensemble
average of which (\ref{eq:c_av_general}) matches the DDFT equation
previously obtained by Rex and Löwen \cite{DDFT_Lowen,DDFT_Hydro_Lowen}.
This generalizes the results of Dean \cite{SPDE_Diffusion_Dean} for
the case of uncorrelated (independent) Brownian walkers to account
for hydrodynamics, and generalizes our prior results \cite{DiffusionJSTAT}
to account for direct interactions among the particles.

A few comments about the physical reasoning behind \eqref{eq:M_R}
are in order. Note that the generic form \eqref{eq:mob_Lowen} does
not fit \eqref{eq:M_R} because the appearance of the Kronecker $\delta_{ij}$.
It can be shown that the requirement that the mobility be positive
semidefinite for any configuration of particles and any $N$ implies
that $\norm{\M{\omega}_{12}(0)}_{2}\leq1$; if this holds as an equality
then %
\footnote{Observe that $\M{\omega}_{12}(0)$ must be rotationally invariant
and therefore has to be the identity matrix.%
} $\M{\omega}_{12}(0)=\M I$ and therefore \eqref{eq:M_R} holds.\textcolor{red}{{}
}This has important physical consequences that do not appear to have
been widely appreciated. Notably, for two overlapping particles, $\V q_{i}=\V q_{j}$,
\eqref{eq:M_R} predicts $\M M_{ii}=\M M_{jj}=\M M_{ij}=\M M_{ji}$,
which implies that, in fact, the two particles continue to move in
synchrony forever, and $\V q_{i}=\V q_{j}$ for all times. By contrast,
if $\norm{\M{\omega}_{12}(0)}_{2}<1$, as for the case of independent
Brownian walkers $\M{\omega}_{12}=0$, two particles released from
the same position separate immediately.

We believe that it is physically more realistic to assume that the
trajectories of nearby particles become highly correlated rather than
remain independent. Furthermore, two perfectly overlapping particles
should behave as if there is only a single particle at that location.
The well-known Rotne-Prager mobility \cite{RotnePrager}, which was
used by Rex and Löwen \cite{DDFT_Lowen,DDFT_Hydro_Lowen} in their
numerical calculations, \emph{does} conform to \eqref{eq:M_R}. In
our prior work \cite{DiffusionJSTAT}, we used a model based on fluctuating
hydrodynamics \cite{LB_SoftMatter_Review,StochasticImmersedBoundary,ForceCoupling_Fluctuations},
which, in the limit of infinite Schmidt number (momentum diffusion
much faster than particle diffusion) converges to (\ref{eq:BD_M})
with \eqref{eq:M_R} and a covariance operator \cite{SIBM_Brownian,BrownianBlobs}
\begin{equation}
\R\left(\V r_{1},\V r_{2}\right)=\frac{k_{B}T}{\eta}\int\M{\sigma}\left(\V r_{1},\V r^{\prime}\right)\M G\left(\V r^{\prime},\V r^{\prime\prime}\right)\M{\sigma}^{T}\left(\V r_{2},\V r^{\prime\prime}\right)d\V r^{\prime}d\V r^{\prime\prime},\label{eq:R_Stokes}
\end{equation}
where $\M G$ is the Green's function for the steady Stokes equation
with unit viscosity and appropriate boundary conditions. For unbounded
three-dimensional systems $\V G$ is the Oseen tensor $\V G\left(\V r^{\prime},\V r^{\prime\prime}\right)=\left(8\pi r\right)^{-1}\left(\M I+r^{-2}\V r\otimes\V r\right)$,
where $\V r=\V r^{\prime}-\V r^{\prime\prime}$. Here $\M{\sigma}$
is a smoothing kernel that filters out features at scales below the
size of the diffusing particles $\sigma$, e.g., $\M{\sigma}$ could
be a Gaussian with standard deviation $\sigma$. The self-diffusion
coefficient $\chi$ given by \eqref{eq:R_Stokes} obeys a Stokes-Einstein
formula, in three dimensions, $\chi\sim k_{B}T/\left(\eta\sigma\right)$
\cite{DiffusionJSTAT}. When the particles are far apart, $\norm{\V q_{i}-\V q_{j}}\gg\sigma$,
the mobility is well-approximated by the Oseen tensor, $\M M_{ij}\left(\V q_{i},\V q_{j}\right)\approx\eta^{-1}\M G\left(\V q_{i},\V q_{j}\right)$.
At short distances the divergence of the Oseen tensor is mollified
by the filter, and (\ref{eq:R_Stokes}) gives a pairwise mobility
very similar to the Rotne-Prager mobility (\ref{eq:RPYTensor}) widely-used
in BD simulations \cite{BrownianBlobs}. Note that (\ref{eq:div_free})
follows from the incompressibility of the Green's function $\M G$.

Numerical methods to solve (\ref{eq:limiting_Ito}) and (\ref{eq:BD_M}),
along with an extensive visual and quantitative analysis of the surprising
characteristics of the solution can be found in Ref. \cite{DiffusionJSTAT}.
A key observation is that, due to the Ito nature of the hydrodynamic
term $-\V w\cdot\grad c$ in (\ref{eq:limiting_Ito}), the ensemble-averaged
concentration continues to follow the local Fick's law (\ref{eq:simple_Fick}),
despite the presence of hydrodynamic correlations among the diffusing
particles. Note, however, that the behavior of each \emph{instance}
(relization) of the stochastic process $c(\V r,t)$ is rather distinct
from the behavior of the mean concentration, as discussed extensively
in our prior work \cite{DiffusionJSTAT}. In particular, the fluctuating
equation (\ref{eq:limiting_Ito}) is non-dissipative (reversible),
while Fick's law (\ref{eq:simple_Fick}) is dissipative (irreversible).
In the presence of large concentration gradients the solutions of
(\ref{eq:limiting_Ito}) exhibit characteristic long-ranged correlations
(giant fluctuations) that are quite distinct from the case of uncorrelated
walkers \cite{GiantFluctConcentration_Sengers,GiantFluctuations_Nature,FractalDiffusion_Microgravity,GiantFluct_NanoColloids}.
This indicates that the mathematical structure and the physical behavior
of (\ref{eq:c_Dean_interacting}) is very different from that of (\ref{eq:c_fluct_general})
because hydrodynamics affects the fluctuations of the concentration
in crucial ways. This fact is well-known in nonequilibrium statistical
mechanics circles, and recent experiments \cite{FractalDiffusion_Microgravity}
have demonstrated how \emph{giant} concentration fluctuations can
arise for a simple polymer solution out of equilibrium in the absence
of gravity. Nonequilibrium concentration fluctuations have also been
measured in gravity for a nanocolloidal suspension \cite{GiantFluct_NanoColloids}.

The striking difference between correlated and uncorrelated walkers
is somewhat surprising. After all, one would expect that, if the correlations
are sufficiently weak in a certain sense (e.g., they decay rapidly
with distance %
\footnote{It must be pointed out, however, that incompressible hydrodynamic
correlations such as the Rotne-Prager tensor, must be long ranged
because of the incompressibility constraint.%
}), (\ref{eq:c_fluct_general}) should converge to (\ref{eq:c_Dean_interacting}).
It is important to emphasize, however, that (\ref{eq:c_Dean_interacting})
corresponds to the physically unrealistic case of particles performing
uncorrelated random motions even when they overlap. In reality, it
is the solvent molecules that have to kick the colloidal particles,
and nearby particles must become correlated because their diffusion
is caused by the motion of correlated fluid molecules. Let us assume
for a moment that (\ref{eq:BD_M}) holds with an isotropic smooth
$\R\left(\V r-\V r^{\prime}\right)$ that is nonzero only if the two
particles are within a distance $\sigma^{\prime}$ apart, and has
a finite value at the origin, $\R\left(0\right)=\chi\M I$. Let us
also account for the fact that the diffusing particles themselves
are not point particles but have a physical size $\sigma$, and consider
the coarse-grained concentration (\ref{eq:c_def_CG}) for $\xi\sim\sigma$.
The case considered by Dean corresponds to the double limit $\sigma\rightarrow0$
and $\sigma^{\prime}\rightarrow0$, but the order of these limits
is not \emph{a priori} clear. Formal manipulations show that (\ref{eq:c_fluct_general})
converges in a certain sense to (\ref{eq:c_Dean_interacting}) if
one takes the limit $\sigma\rightarrow0$ first and then takes the
limit $\sigma^{\prime}\rightarrow0$. It is an interesting open question
what happens if the order of the limits is reversed, or if $\sigma$
and $\sigma^{\prime}$ go to zero simultaneously. Such calculations
will shed further light on the nature of diffusion in liquid suspensions
and mixtures over a much broader spectrum of length and time scales
than described by Fick's law with phenomenological diffusion constants.

\rule[0.5ex]{0.75\columnwidth}{1pt}
\begin{acknowledgments}
We are grateful to Pep Espanol and Mike Cates for their insightful
comments.\emph{ }A. Donev was supported in part by the National Science
Foundation under grant DMS-1115341 and the Office of Science of the
U.S. Department of Energy through Early Career award DE-SC0008271.
E. Vanden-Eijnden was supported by the DOE office of Advanced Scientific
Computing Research under grant DE-FG02-88ER25053, by the NSF under
grant DMS07-08140, and by the Office of Naval Research under grant
N00014-11-1-0345.
\end{acknowledgments}
\appendix

\section*{Appendix}

\section{\label{sec:Derivations}Equations for the Empirical Concentration}

In this Appendix we present the detailed derivation of (\ref{eq:c_fluct_general})
and (\ref{eq:c_av_general}). In the beginning, we will consider the
case of no direct interactions among the particles, $U=0$, and subsequently
add the direct forces. The assumption that the covariance operator
$\R$ is symmetric positive-semidefinite is equivalent to the requirement
that the mobility matrix $\M M\left(\V Q\right)$ be symmetric positive
semi-definite for all $\V Q$, and implies that there exists an infinite
countable set of eigenfunctions $\V{\phi}_{k}(\V r)$ that factorize
(diagonalize) the covariance operator,
\[
\sum_{k}\V{\phi}_{k}(\V r)\otimes\V{\phi}_{k}(\V r^{\prime})=\R(\V r,\V r^{\prime}).
\]
Note that if (\ref{eq:div_free}) holds then the eigenfunctions of
$\R$ are also incompressible, $\grad\cdot\V{\phi}_{k}(\V r)=0$.

\subsection{Stratonovich form}

It is not hard to show that in the absence of direct interactions
(\ref{eq:BD_M},\ref{eq:M_R}) corresponds to the Stratonovich equation
for the position of an individual tracer $i=1,\ldots,N$, 
\begin{equation}
\begin{aligned}d\V q_{i} & =\sum_{j=1}^{N}\V b(\V q_{i},\V q_{j})dt+\sqrt{2}\sum_{k}\V{\phi}_{k}(\V q_{i})\circ dB_{k},\end{aligned}
\label{eq:1S}
\end{equation}
where $\circ$ denotes a Stratonovich product, $B_{k}$ are independent
Brownian motions, and we defined 
\begin{equation}
\V b(\V r,\V r^{\prime})=\grad^{\prime}\cdot\R(\V r,\V r^{\prime})=\sum_{k}\V{\phi}_{k}(\V r)\grad\cdot\V{\phi}_{k}(\V r^{\prime}).\label{eq:12}
\end{equation}
Note that when the incompressibility condition (\ref{eq:div_free})
holds $\V b(\V r,\V r^{\prime})=0.$

For the Stratonovich interpretation we can use ordinary calculus to
write 
\begin{equation}
\begin{aligned}dc(\V r,t) & =-\sum_{i,j=1}^{N}\V b(\V q_{i}(t),\V q_{j}(t))\cdot\grad\delta(\V r-\V q_{i}(t))dt\\
 & \quad-\sqrt{2}\sum_{i=1}^{N}\sum_{k}\V{\phi}_{k}(\V q_{i}(t))\cdot\grad\delta(\V r-\V q_{i}(t))\circ dB_{k}(t)
\end{aligned}
\label{eq:6S}
\end{equation}
Using integration by parts and properties of the delta function we
can write this as a closed-form equation for $c$, 
\begin{equation}
\begin{aligned}dc(\V r,t) & =-\grad\cdot\left(c(\V r,t)\int\V b(\V r,\V r^{\prime})c(\V r^{\prime},t)d\V r^{\prime}\right)dt\\
 & \quad-\sqrt{2}\sum_{k}\grad\cdot\left(\V{\phi}_{k}(\V r)c(\V r,t)\right)\circ dB_{k}(t)
\end{aligned}
\label{eq:6bS}
\end{equation}
or, after recalling the definition of $\V b$ in (\ref{eq:12}) and
performing an integration by parts to transfer the gradient to $c$,
\begin{equation}
\begin{aligned}dc(\V r,t) & =\grad\cdot\left(c(\V r,t)\int\R(\V r,\V r^{\prime})\grad^{\prime}c(\V r^{\prime},t)d\V r^{\prime}\right)dt\\
 & \quad-\sqrt{2}\sum_{k}\grad\cdot\left(\V{\phi}_{k}(\V r)c(\V r,t)\right)\circ dB_{k}(t).
\end{aligned}
\label{eq:6cS}
\end{equation}
When the incompressibility condition (\ref{eq:div_free}) is satisfied,
$\V b=0$ and (\ref{eq:6bS}) implies that 
\begin{equation}
\begin{aligned}dc(\V r,t)=-\sqrt{2}\sum_{k}\V{\phi}_{k}(\V r)\cdot\grad c(\V r,t)\circ dB_{k}(t)\end{aligned}
,\label{eq:6incS}
\end{equation}
which is exactly identical to the Stratonovich form of the equation
we obtained in Ref. \cite{DiffusionJSTAT} by rather different means.
Here we can identify $\V w\left(\V r,t\right)=\sqrt{2}\sum_{k}\V{\phi}_{k}(\V r)dB_{k}(t)$
as a random velocity field with covariance given by (\ref{eq:C_w}).
While the Stratonovich form of the equation is the simplest, the Ito
form is the most convenient for performing an ensemble average to
obtain an equation for the average concentration $c^{(1)}$.

\subsection{Ito form}

In the Ito interpretation, (\ref{eq:1S}) reads 
\begin{equation}
d\V q_{i}=\V a(\V q_{i})dt+\sum_{j\not=i}^{N}\V b(\V q_{i},\V q_{j})dt+\sqrt{2}\sum_{k}\V{\phi}_{k}(\V q_{i})dB_{k},\label{eq:1}
\end{equation}
where we defined 
\begin{equation}
\begin{aligned}\V a(\V r) & =\grad\cdot\R(\V r,\V r)=\sum_{k}\V{\phi}_{k}(\V r)\grad\cdot\V{\phi}_{k}(\V r)+\sum_{k}\V{\phi}_{k}(\V r)\cdot\grad\V{\phi}_{k}(\V r)=\V b(\V r,\V r)+\V g(\V r),\end{aligned}
\label{eq:12-1}
\end{equation}
and
\begin{equation}
\V g(\V r)=\sum_{k}\V{\phi}_{k}(\V r)\cdot\grad\V{\phi}_{k}(\V r).\label{eq:19}
\end{equation}
The Ito equation \eqref{eq:1} can also be written as 
\begin{equation}
d\V q_{i}=\V g(\V q_{i})dt+\sum_{j=1}^{N}\V b(\V q_{i},\V q_{j})dt+\sqrt{2}\sum_{k}\V{\phi}_{k}(\V q_{i})dB_{k},\label{eq:1new}
\end{equation}
which will be the most convenient for our calculation. Note that when
the incompressibility condition (\ref{eq:div_free}) holds, 
\begin{equation}
\V a(\V r)=\V g(\V r)=\grad\cdot\M{\chi}\left(\V r\right)\label{eq:20}
\end{equation}
is the divergence of the diffusion tensor, which vanishes for translationally-invariant
systems.

Using Ito calculus, we can now write an equation for the empirical
concentration, 
\begin{equation}
\begin{aligned}dc(\V r,t) & =-\sum_{i=1}^{N}\left(\V g(\V q_{i}(t))+\sum_{j=1}^{N}\V b(\V q_{i}(t),\V q_{j}(t))\right)\cdot\grad\delta(\V r-\V q_{i}(t))dt\\
 & \quad-\sqrt{2}\sum_{i=1}^{N}\sum_{k}\V{\phi}_{k}(\V q_{i}(t))\cdot\grad\delta(\V r-\V q_{i}(t))dB_{k}(t)\\
 & \quad+\sum_{i=1}^{N}\sum_{k}\V{\phi}_{k}(\V q_{i}(t))\V{\phi}_{k}(\V q_{i}(t)):\grad\grad\delta(\V r-\V q_{i}(t))dt
\end{aligned}
\label{eq:6}
\end{equation}
Using integration by parts and properties of the delta function we
can write this as a closed-form equation for $c$, 
\begin{equation}
\begin{aligned}dc(\V r,t) & =-\grad\cdot\left(\V g(\V r)c(\V r,t)\right)dt+\grad\grad:\left(\R(\V r,\V r)c(\V r,t)\right)dt\\
 & \quad-\grad\cdot\left(c(\V r,t)\int\V b(\V r,\V r^{\prime})c(\V r^{\prime},t)d\V r^{\prime}\right)dt\\
 & \quad-\sqrt{2}\sum_{k}\grad\cdot\left(\V{\phi}_{k}(\V r)c(\V r,t)\right)dB_{k}(t),
\end{aligned}
\label{eq:6b}
\end{equation}
which can further be simplified to 
\begin{equation}
\begin{aligned}dc(\V r,t) & =\grad\cdot\left(\R(\V r,\V r)\grad c(\V r,t)+\V b(\V r,\V r)c(\V r,t)\right)dt\\
 & +\grad\cdot\left(c(\V r,t)\int\R(\V r,\V r^{\prime})\grad^{\prime}c(\V r^{\prime},t)d\V r^{\prime}\right)dt\\
 & -\sqrt{2}\sum_{k}\grad\cdot\left(\V{\phi}_{k}(\V r)c(\V r,t)\right)dB_{k}(t).
\end{aligned}
\label{eq:6c}
\end{equation}
Here we can identify $\V w\left(\V r,t\right)=\sqrt{2}\sum_{k}\V{\phi}_{k}(\V r)dB_{k}(t)$
as a random velocity field with covariance given by (\ref{eq:C_w}).

Upon averaging over realizations of the noise the fluctuating term
drops out in the Ito interpretation, giving a non-local diffusion
equation for the mean concentration $c^{(1)}(\V r,t)=\av{c(\V r,t)}$
\begin{eqnarray}
\partial_{t}c^{(1)}(\V r,t) & = & \grad\cdot\left(\M{\chi}(\V r)\grad c^{(1)}(\V r,t)+\V b(\V r,\V r)c^{(1)}(\V r,t)\right)\nonumber \\
 & + & \grad\cdot\left(\int\R(\V r,\V r^{\prime})\av{c\left(\V r,t\right)\grad^{\prime}c\left(\V r^{\prime},t\right)}d\V r^{\prime}\right).\label{eq:dc_dt_mean}
\end{eqnarray}
By noting that the two-particle correlation function is
\begin{equation}
c^{(2)}\left(\V r,\V r^{\prime},t\right)=\av{c\left(\V r,t\right)c\left(\V r^{\prime},t\right)}-\av{c\left(\V r,t\right)}\delta\left(\V r-\V r^{\prime}\right),\label{eq:c2_def}
\end{equation}
we see after an integration by parts that (\ref{eq:dc_dt_mean}) is
equivalent to
\[
\partial_{t}c^{(1)}(\V r,t)=\grad\cdot\left(\M{\chi}(\V r)\grad c^{(1)}(\V r,t)\right)+\grad\cdot\left(\int\R(\V r,\V r^{\prime})\grad^{\prime}c^{(2)}\left(\V r,\V r^{\prime},t\right)\, d\V r^{\prime}\right),
\]
which, for a translationally-invariant system, is exactly the equation
(\ref{eq:c_t_Lowen}) obtained by Rex and Löwen.

When the incompressibility condition (\ref{eq:div_free}) holds, Eq.
(\ref{eq:6b}) reduces to the fluctuating Fick's law 
\begin{equation}
\begin{aligned}dc(\V r,t) & =\grad\cdot\left(\M{\chi}(\V r)\grad c(\V r,t)\right)dt-\sqrt{2}\sum_{k}\V{\phi}_{k}(\V r)\cdot\grad c(\V r,t)dB_{k}(t),\end{aligned}
\label{eq:6inc}
\end{equation}
which is exactly the stochastic advection-diffusion equation (\ref{eq:limiting_Ito}).
In this case the mean follows Fick's law
\[
\partial_{t}c^{(1)}(\V r,t)=\grad\cdot\left(\M{\chi}(\V r)\grad c^{(1)}(\V r,t)\right),
\]
and the non-local diffusion term involving $c^{(2)}$ disappears since
$\V b=0$.

\subsection{Direct interactions}

If we include direct interactions among the particles of the form
(\ref{eq:U_Q}), the Stratonovich equation of motion \eqref{eq:1S}
becomes 
\begin{equation}
\begin{aligned}d\V q_{i} & =\sum_{j=1}^{N}\V b(\V q_{i},\V q_{j})dt+\sqrt{2}\sum_{k}\V{\phi}_{k}(\V q_{i})\circ dB_{k}\\
 & +\left(k_{B}T\right)^{-1}\sum_{j=1}^{N}\R(\V q_{i},\V q_{j})\left(\V f_{1}(\V q_{j})+\sum_{k\not=j}\V f_{2}(\V q_{j},\V q_{k})\right)dt
\end{aligned}
\label{eq:1Sf}
\end{equation}
where we have defined the external and pairwise forces 
\begin{equation}
\V f_{1}(\V r)=-\grad U_{1}(\V r),\qquad\V f_{2}(\V r,\V r^{\prime})=-\grad U_{2}(\V r,\V r^{\prime}).\label{eq:4}
\end{equation}
For simplicity, and without loss of generality, we will assume that
there is no self-force coming from the pairwise interactions, $\V f_{2}(\V r,\V r)=0$.

The new term in~\eqref{eq:1Sf} adds the following term to the drift
in~\eqref{eq:6S}: 
\begin{equation}
\begin{aligned} & -\left(k_{B}T\right)^{-1}\sum_{i,j=1}^{N}\R(\V q_{i},\V q_{j})\left(\V f_{1}(\V q_{j})+\sum_{k\not=j}\V f_{2}(\V q_{j},\V q_{k})\right)\cdot\grad\delta(\V r-\V q_{i})\\
 & =-\left(k_{B}T\right)^{-1}\grad\cdot\left[\sum_{i,j=1}^{N}\R(\V q_{i},\V q_{j})\left(\V f_{1}(\V q_{j})+\sum_{k=1}^{N}\V f_{2}(\V q_{j},\V q_{k})\right)\delta(\V r-\V q_{i})\right]\\
 & =-\left(k_{B}T\right)^{-1}\grad\cdot\left(\sum_{i,j=1}^{N}\int\R(\V r,\V r^{\prime})\V f_{1}(\V r^{\prime})\delta(\V r-\V q_{i})\delta(\V r^{\prime}-\V q_{j})\, d\V r^{\prime}\right)\\
 & \quad-\left(k_{B}T\right)^{-1}\grad\cdot\left(\sum_{i,j,k=1}^{N}\int\R(\V r,\V r^{\prime})\V f_{2}(\V r^{\prime},\V r^{\dprime})\delta(\V r-\V q_{i})\delta(\V r^{\prime}-\V q_{j})\delta(\V r^{\dprime}-\V q_{k})\, d\V r^{\prime}d\V r^{\dprime}\right)
\end{aligned}
\label{eq:5}
\end{equation}
This can also be written in terms of the empirical concentration as
\begin{equation}
\begin{aligned} & -\left(k_{B}T\right)^{-1}\grad\cdot\left(c(\V r,t)\int\R(\V r,\V r^{\prime})\V f_{1}(\V r^{\prime})c(\V r^{\prime},t)\, d\V r^{\prime}\right)\\
 & -\left(k_{B}T\right)^{-1}\grad\cdot\left(c(\V r,t)\int\R(\V r,\V r^{\prime})\V f_{2}(\V r^{\prime},\V r^{\dprime})c(\V r^{\prime},t)c(\V r^{\dprime},t)\, d\V r^{\prime}d\V r^{\dprime}\right),
\end{aligned}
\label{eq:7}
\end{equation}
or, in terms of the potentials $U_{1}$ and $U_{2}$, as 
\begin{equation}
\begin{aligned} & \left(k_{B}T\right)^{-1}\grad\cdot\left(c(\V r,t)\int\R(\V r,\V r^{\prime})\grad^{\prime}U_{1}(\V r^{\prime})c(\V r^{\prime},t)\, d\V r^{\prime}\right)\\
+ & \left(k_{B}T\right)^{-1}\grad\cdot\left(c(\V r,t)\int\R(\V r,\V r^{\prime})\grad^{\prime}U_{2}(\V r^{\prime},\V r^{\dprime})c(\V r^{\prime},t)c(\V r^{\dprime},t)\, d\V r^{\prime}d\V r^{\dprime}\right).
\end{aligned}
\label{eq:7pot}
\end{equation}
Adding these terms to the right hand side of (\ref{eq:6c}) (or, in
the Stratonovich interpretation, to \eqref{eq:6cS}) gives our final
result (\ref{eq:c_fluct_general}). Remarkably, this is a closed equation
for the fluctuating concentration just as in the case of uncorrelated
particles \cite{SPDE_Diffusion_Dean}.

Taking an ensemble average of the new terms (\ref{eq:7pot}) leads
to terms involving the two-particle (\ref{eq:c2_def}) and three-particle
correlation function
\begin{equation}
\begin{aligned}c^{(3)}(\V r,\V r^{\prime},\V r^{\dprime},t) & =\langle c(\V r,t)c(\V r^{\prime},t)c(\V r^{\dprime},t)\rangle\\
 & -\tfrac{1}{2}\langle c(\V r,t)c(\V r^{\prime},t)\rangle\left(\delta(\V r-\V r^{\dprime})+\delta(\V r^{\prime}-\V r^{\dprime})\right)\\
 & -\tfrac{1}{2}\langle c(\V r,t)c(\V r^{\dprime},t)\rangle\left(\delta(\V r-\V r^{\prime})+\delta(\V r^{\prime}-\V r^{\dprime})\right)\\
 & -\tfrac{1}{2}\langle c(\V r^{\prime},t)c(\V r^{\dprime},t)\rangle\left(\delta(\V r-\V r^{\prime})+\delta(\V r-\V r^{\dprime})\right)\\
 & +\langle c(\V r,t)\rangle\delta(\V r-\V r^{\prime})\delta(\V r^{\prime}-\V r^{\dprime}),
\end{aligned}
\label{eq:22c}
\end{equation}
which, after some algebra, gives our final result (\ref{eq:c_av_general})
for the average concentration.


\end{document}